\documentstyle[times,10pt,epsfig,amssymb,nato]{crckapb}

\newcommand{\HII}{\mbox{H\,{\footnotesize II}}}       
\newcommand{\HI}{\mbox{H\,{\footnotesize I}}}         
\newcommand{\phn}{\phantom{0}}

\begin{opening}

\title{OB Associations}

\author{A.G.A.~Brown}
\institute{Instituto de Astronom{\'\i}a UNAM,\\
  Apartado Postal 877, Ensenada, 22800 Baja California, Mexico}

\author{A.~Blaauw}
\institute{Kapteyn Instituut, Postbus 800, 9700 AV Groningen, The Netherlands\\
  Sterrewacht Leiden, Postbus 9513, 2300 RA Leiden, The Netherlands}

\author{R.~Hoogerwerf}
\author{J.H.J.~de Bruijne}
\author{P.T.~de Zeeuw} 
\institute{Sterrewacht Leiden, Postbus 9513, 2300 RA Leiden, The Netherlands}

\end{opening}

\begin{document}

%
%

\section{Introduction}\label{sec:introduction}

Starting with the earliest studies of the distribution of the bright stars by
Kapteyn, Rasmuson, and Pannekoek (\cite{Kapteyn14}, \cite{Kapteyn18},
\cite{Pannekoek29}, \cite{Rasmuson21}, and \cite{Rasmuson27}), it was evident
that O and B stars are not distributed randomly on the sky, but instead are
concentrated in loose groups, which were subsequently called `OB
associations'. This inspired research on their individual properties, and on
their motions and space distribution. From dynamical considerations it
followed that OB associations must be young \cite{Ambartsumian49}, a
conclusion supported later by the ages derived from color-magnitude diagrams.
Blaauw's 1964 review \cite{Blaauw64} already discussed the relation between OB
associations and interstellar matter. Subsequent observations of molecular
clouds (e.g., \cite{Zuckerman74}, \cite{Blitz80}) indicated that these groups
are usually located in or near starforming regions, and hence are prime sites
for the study of star formation processes and of the interaction of early-type
stars with the interstellar medium.

An important aspect of the study of OB associations is the identification of
their `members'. Associations have small internal velocity dispersions (e.g.,
\cite{Mathieu86}, \cite{Tian96}), so that the streaming motion of the
association as a whole, as well as the Solar motion, is reflected as a motion
of the members towards a convergent point on the sky (e.g., \cite{Bertiau58}).
This can be used to establish membership of these `moving groups' based on
measurements of proper motions. At the time of the previous NATO-ASI on star
formation \cite{CreteI} few astrometric membership studies existed for nearby
OB associations because these generally cover tens to hundreds of square
degrees on the sky.  Ground-based proper motion studies therefore almost
invariably had been confined to modest samples of bright stars ($V\lesssim 6$)
in fundamental or meridian circle catalogues, or to small areas covered by a
single photographic plate. Photometric studies can extend membership to later
spectral types (e.g., \cite{Warren77a}--\cite{Warren78}), but are less
reliable due to, e.g., undetected duplicity, or the distance spread within an
association. As a result, membership for many associations had been determined
unambiguously only for spectral types earlier than B5 (e.g., \cite{Blaauw64},
\cite{Blaauw91}).

\begin{figure}[tbh]
  \begin{center}
    \epsfig{file=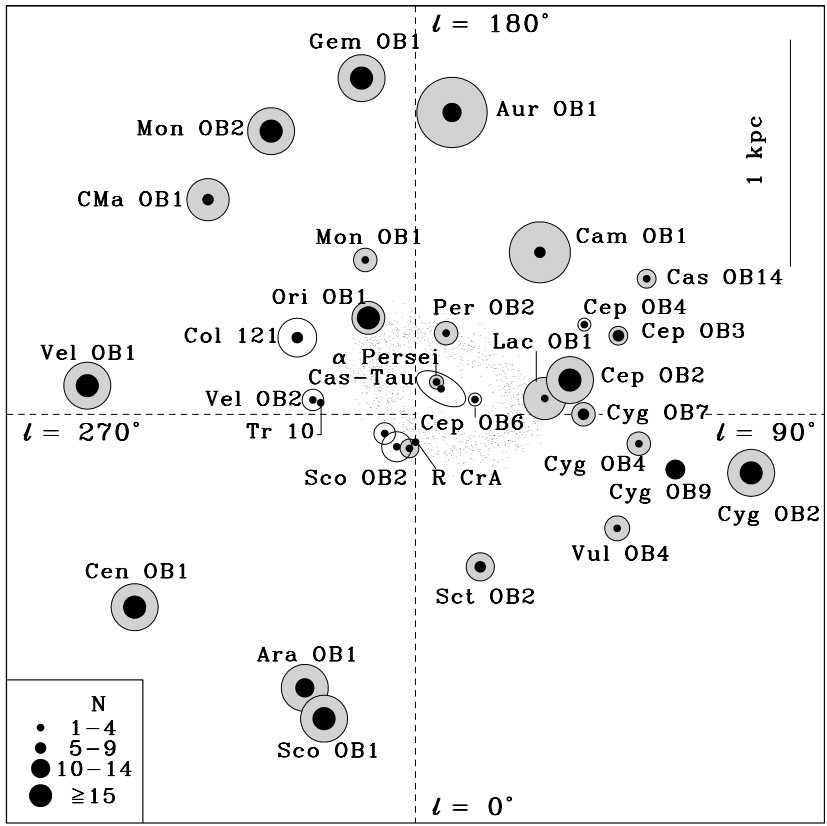,height=11truecm}
  \end{center}
  \caption{Locations of the OB associations within $\sim$1.5~kpc,
    projected onto the Galactic plane. This figure is an update of Figure~8 in
    \protect\cite{Blaauw91} and is based on a list by Ruprecht
    \protect\cite{Ruprecht66} (see de Zeeuw et~al.\ \protect\cite{Z99} for
    more details). The Sun is at the center of the dashed lines which give the
    principal directions in Galactic longitude, $\ell$. The sizes of the
    circles represent the projected dimensions of the associations, enlarged
    by a factor 2 with respect to the distance scale. The sizes of the central
    dots indicate the degree of current or recent star formation activity, as
    given by the number $N$ of stars more luminous than absolute magnitude
    $M_V\sim$$-5$ \protect\cite{Humphreys78}.  Associations which are absent
    from Ruprecht's list are represented as open circles. The distribution of
    small dots indicates the Gould Belt \protect\cite{Poppel97}.
    Figure~\ref{fig:gould} presents the post-Hipparcos map of the nearby OB
    associations.}
  \label{fig:ruprecht}
\end{figure}

This review is concentrated on OB associations located within $\sim 1.5$~kpc
from the Sun, and updates and extends the chapter by Blaauw in the previous
edition of this NATO-ASI \cite{Blaauw91}. The material discussed in Blaauw's
chapter is still relevant but we chose to emphasize the developments in the
field since 1991. The X-ray surveys by the EINSTEIN and ROSAT satellites
enabled the identification of large numbers of low-mass stars in OB
associations. Major progress in the identification of high-mass members of OB
associations has been made possible through the use of precise astrometric data
from the Hipparcos Catalogue \cite{ESA97}.  Also, Blaauw's 1964 review
\cite{Blaauw64} still is essential reading. Reviews concerning OB associations
throughout the Local Group include \cite{Garmany94}, \cite{Massey98a}, and
\cite{Melnick92}.  In addition there are conference proceedings that contain
material on OB associations: \cite{Boland85}, \cite{Cassinelli93},
\cite{Janes91}, \cite{McCaughrean98}, and \cite{TT92}. For more background on
the historical development of OB association research we refer to
\cite{Blaauw64}, \cite{Brown98}, \cite{Elmegreen98}, and \cite{Z99}.

This chapter proceeds as follows. \S \ref{sec:why} discusses the reasons for
studying OB associations. \S \ref{sec:define} describes the problem of
accurately defining what is meant by an OB association. \S \ref{sec:recent}
contains an overview of recent ground-based studies of OB associations, and \S
\ref{sec:census} describes the Hipparcos census of the nearby OB associations
\cite{Z99}. \S \ref{sec:gould} contains a discussion of Gould's Belt and the
origins of OB associations.  \S \ref{sec:future} points out directions for
future research.

Most of the OB associations discussed here appear in a list compiled by
Ruprecht \cite{Ruprecht66}, which contains field boundaries, some bright
members, distances, and uses a consistent nomenclature, which was subsequently
approved by the IAU.  Figure~\ref{fig:ruprecht} shows the distribution of the
nearby OB associations, based on Ruprecht's list and including additional
associations discussed by De Zeeuw et~al.\ \cite{Z99}, projected onto the
Galactic plane.

\section{The Importance of Studying OB Associations}\label{sec:why}

In the context of star formation a detailed examination of the stellar
content, structure, and kinematics of OB associations allows us to address the
following fundamental questions:

\begin{itemize}

\item[$\bullet$] What is the initial mass function? Young stellar groups are
  prime sites for the study of the initial mass function (IMF) because the
  corrections for stellar evolution and star formation history are minimal or
  at least straightforward \cite{Scalo98}. Moreover, OB associations contain
  the entire range of stellar masses (cf.\ \S\S \ref{sec:highmass}
  and~\ref{sec:lowmass}).

\item[$\bullet$] What are the characteristics of the binary population?  A
  detailed answer to this question will furnish a better understanding of the
  processes through which binaries form, but also of the formation of massive
  stars and star clusters. Recent theories (\cite{Bonnell97},
  \cite{Bonnell98}, and the chapter by Bonnell in this volume) predict the
  occurrence of few widely separated binaries among systems with massive
  primaries. A characterization of the binary population in OB associations is
  a direct test of these theories.

\item[$\bullet$] OB associations have traditionally been characterized as
  unbound groups of stars \cite{Blaauw64}. However, the youngest subgroup of
  Ori~OB1, the Orion Nebula Cluster, may eventually evolve into a bound
  cluster \cite{Hillenbrand98}. What causes the distinction between the
  formation of bound open clusters and unbound associations? Various studies
  show that the fate of a particular young group of stars depends on the gas
  fraction, the time scale on which this gas is removed and the stellar
  velocity dispersion (e.g., \cite{CLada84}, \cite{Verschueren89}). However,
  the processes that set the values of the above parameters during the
  formation of a star cluster are not well understood.

\item[$\bullet$] Observations show that molecular clouds and cloud cores
  contain much more angular momentum than the stars that form from them (e.g.,
  \cite{Bodenheimer93}, \cite{Mouschovias91}). How is angular momentum
  redistributed during star formation? What is the resulting distribution of
  rotational velocities of stars?  Important clues to the answers of these
  questions may be obtained from studies of the distribution of rotational
  velocities of the members of OB associations in conjunction with studies of
  their binary population.

\item[$\bullet$] O- and B-type runaway stars are a subset of the O and B stars
  characterized by their high space velocities, up to 200~km~s$^{-1}$, and
  almost complete absence of multiplicity. For a number of these stars, the
  motion through space, when traced backwards in time, leads to the
  identification of a parent OB association. Possible mechanisms for the
  origin of the OB runaways are: release of the secondary following the
  supernova explosion of the primary member of a binary \cite{Blaauw61}, and
  purely dynamical interactions amongst members of a protocluster
  \cite{Gies86}. The precise characterization of OB runaways as well as more
  identifications of parent associations can provide more insight into massive
  binary evolution \cite{Rensbergen96} and massive star formation
  \cite{Clarke92}. OB runaways were most recently reviewed by Blaauw
  \cite{Blaauw93}.

\end{itemize}

Further questions related to star formation that can be addressed through
studies of OB associations include the star formation rate and efficiency, the
issue of the star formation history in clusters and associations (i.e., do all
stars in a group form at the same time?), and the propagation of star
formation throughout molecular cloud complexes (i.e., sequential star
formation \cite{Elmegreen77}). In a broader context the investigation of OB
associations is important for numerous areas of Galactic and extragalactic
research. A few examples follow:

\begin{itemize}

\item[$\bullet$] The calibration of the absolute luminosity of the upper main
  sequence is based largely on the bright members of the nearest OB
  associations (e.g.,~\cite{Blaauw56}).

\item[$\bullet$] Massive stars influence the interstellar medium through their
  ionizing radiation, stellar winds, and supernovae. They may be responsible
  for determining the pressure of the gas and the velocity dispersion of
  atomic clouds \cite{McKee77}, and for the destruction of the molecular
  clouds in which they are born. Acting collectively, massive stars in OB
  associations have an even larger impact on the evolution of the interstellar
  medium and create \HI{} supershells and superbubbles (e.g., \cite{TT88}).

\item[$\bullet$] The recent discovery of unexpectedly large amounts of
  gamma-ray emission in the direction of the Orion molecular cloud complex
  \cite{Bloemen94} have been linked to the presence of the nearby O and B
  stars in Ori~OB1 (e.g., \cite{Parizot98}).

\end{itemize}

The nearby associations contain few massive stars: even in Ori~OB1 the upper
mass limit is only about 50~M$_\odot$. More distant associations enable the
determination of the IMF for the most massive stars. A well-known example of
an OB association beyond $1.5$~kpc is NGC 3603, which contains 50 O stars and
ionizes the second most luminous \HII\ region in the Galaxy
\cite{Eisenhauer98}. The Carina \HII\ region/ molecular cloud complex contains
numerous star clusters, the most famous of which are Trumpler~14 and 16
\cite{Brooks98}.  The latter are among the most massive clusters in the Galaxy
and contain many O stars, including some of type O3 ($\gtrsim 80$~M$_\odot$).

A better characterization of the IMF at the high-mass end, the luminosities
and spectral characteristics of massive stars, and the fraction of Wolf--Rayet
stars in OB associations are essential items to consider when interpreting
observations of extragalactic starforming regions and starburst galaxies.

\section{Defining Characteristics of OB Associations}
\label{sec:define}

In 1947 Ambartsumian \cite{Ambartsumian47} introduced the term `association'
for groups of OB stars; he pointed out that their stellar mass density is
usually less than 0.1 ${\rm M}_\odot$~pc$^{-3}$. Bok \cite{Bok34} had already
shown that such low-density stellar groups are unstable against Galactic tidal
forces, which led Ambartsumian to conclude that OB associations must be young
($\sim$10~Myr) \cite{Ambartsumian49}. Because of these low stellar densities
associations will quickly disperse. This makes them hard to recognize once
they are older than about 25~Myr due to their large spatial extent
($\sim$10--50~pc). These properties were used in the earliest definitions of
OB associations given by Ambartsumian and Blaauw \cite{Ambartsumian47},
\cite{Blaauw64}. However, such definitions do not necessarily encompass all
associations. For example, the Orion Nebula Cluster is considered a subgroup
of the Ori~OB1 association, but is likely to remain bound and thus compact
\cite{Hillenbrand98}. If the cluster in reality is unbound a velocity
dispersion of a few km~s$^{-1}$ will be enough to disperse it from its present
1~pc size to a size of a typical OB association within 5--10~Myr.

Another definition of an OB association was proposed by Lada \& Lada
\cite{CLada91}; a group of at least 10 members whose stellar density is less
than 1~M$_\odot$~pc$^{-3}$. However, as the authors themselves pointed out,
this would mean that the Hyades would be classified as an OB association and
the association NGC~2264 as an open cluster. This led Brown et~al.\ 
\cite{Brown97a} to define OB associations as those stellar groups that are
left unbound after the process of gas removal from the protocluster has been
concluded. This is not a very practical definition as it is not easy to
determine whether a particular stellar group is bound or not.

Recently, the problem of accurately defining the term `OB association' has
been discussed by Elmegreen \& Efremov \cite{Elmegreen98}.  Numerous
observational selection effects that occur when looking for OB associations
are considered. For instance, how stellar groups are selected depends on
distance, because a particular group of luminous stars can resemble a dense
cluster if it is at a large distance, or a rarefied association if it is
nearby.  Observations of patches of Cepheid variables in the LMC and in local
star formation regions \cite{Elmegreen96} reveal a correlation between the
duration of star formation in a region and its size. This means that if an OB
association is defined as a grouping of OB and other young stars of, for
example, less than 10~Myr, the association will be observed to have a maximum
size of about 50~pc because larger regions will generally contain older stars.
Elmegreen \& Efremov argue that these difficulties arise because there is no
physical scale associated with star formation.  Rather, there is a whole
hierarchy of scales varying from single/binary stars to star clusters, star
complexes, and even up to small pieces of a spiral arm.  In combination with
the size-age correlation mentioned above this will lead to stellar groupings
with particular dimensions being selected according to the age limits imposed.
Hence, if one allows for enough range in age it could be argued that all of
Gould's Belt, consisting of several associations and a wider distribution of
early-type stars in the Solar neighbourhood, is actually one association. It
is further argued \cite{Elmegreen98} that the hierarchical and scale-free
nature of star formation results from the fractal structure of interstellar
gas. However, McKee \& Williams (\cite{McKee97}, \cite{Williams97}) showed
that there is a physical upper limit to the size of OB associations and giant
molecular clouds. This finding supports the idea that the characteristic sizes
observed for OB associations ($\sim$$10$--50~pc) are a reflection of the sizes
of giant molecular clouds from which they form.  Moreover, it was recently
inferred that the nearest molecular cloud complex, in Taurus, has a
characteristic size scale and hence does not have a fractal structure
\cite{Blitz97}.

Based on studies of Hertzsprung--Russell diagrams for OB associations it was
realized early on \cite{Blaauw64} that several associations could be divided
further into subgroups. These can primarily be distinguished on the basis of
the ages of their members and their degree of association with interstellar
matter (\cite{Blaauw64}, \cite{DeGeus89}, \cite{Warren77a}, \cite{Warren78}).
Defining the boundaries of these subgroups is a thorny question. To date, one
cannot distinguish subgroups kinematically, not even with proper motions of
Hipparcos accuracy (see \S \ref{sec:census}). A good example is Ori~OB1.
Originally this association was divided into four subgroups 1a--1d
\cite{Blaauw64}. Subgroup 1b consists of the stars around Orion's Belt and was
further subdivided in some studies, but was recently treated as a whole
\cite{Brown94}. The latest finding on the Belt stars is that the star $\sigma$
Ori is surrounded by a cluster of young stars, which may represent an older
analogue of the Trapezium cluster \cite{Walter97}, \cite{Walter98}. This
suggests that more such clusters of pre-main sequence stars surrounding bright
OB stars may be found and a further subdivision of subgroup 1b may thus be
warranted.  This serves as a warning that established subgroup boundaries in
associations are not `cast in stone' and that future work may lead to
different subdivisions.

In summary, it is not entirely trivial to accurately define what is meant by
the term `OB association'. Nevertheless, for the purpose of this review we
define OB associations to be young ($\lesssim$$50$~Myr) stellar groupings of
low density---such that they are likely to be unbound---containing a
significant population of B stars. Their projected dimensions range from
$\sim$10 to $\sim$100~pc.

\section{Recent Research}\label{sec:recent}

We now turn to work on OB associations carried out since 1991 but before the
availability of Hipparcos data, and grouped by subject. We first concentrate
on the high-mass stars and then discuss low-mass stars. This overview is not
intended to be complete but serves as an introduction to recent literature on
OB associations.

\subsection{High-Mass Stars}\label{sec:highmass}

\paragraph{Stellar Content.}
Extensive studies of the stellar content of OB associations were carried out
using photometric data. As explained in \S \ref{sec:introduction} this allowed
an extension of membership lists to later spectral types, for which no
accurate large-scale proper motion surveys were available yet. The work on
Sco~OB2 was summarized in \cite{Blaauw91} and the details can be found in
\cite{DeGeus89}.

\begin{figure}[tbh]
  \begin{center}
    \epsfig{file=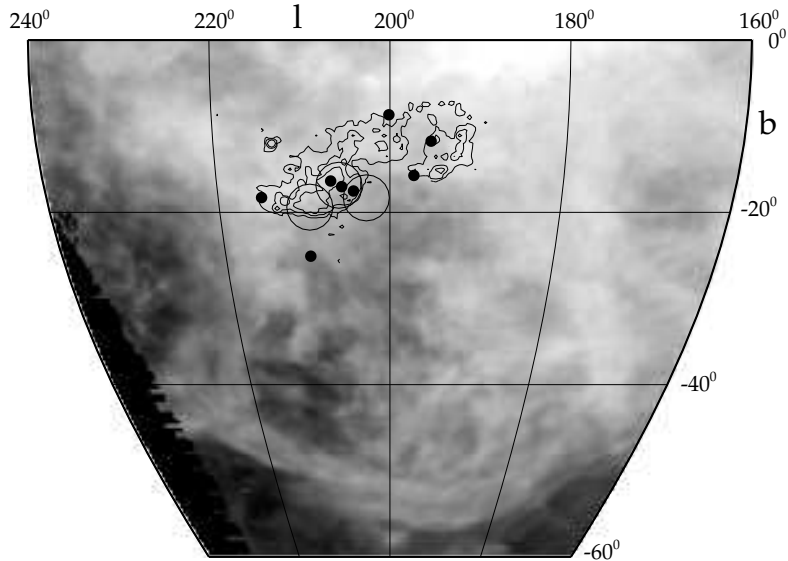,height=7.5truecm}
  \end{center}
  \caption{The Orion--Eridanus bubble. The grey-scale image is a
    logarithmically scaled representation of integrated \mbox{H\,{\tiny I}}
    emission from the Leiden--Dwingeloo survey in the velocity interval
    $-1$~km~s$^{-1}\le v_{\tiny\textrm{LSR}}\le +8$~km~s$^{-1}$. The contours
    outline the 100~$\mu$m (IRAS) emission from the Orion A and B molecular
    clouds (the ring around $(\ell,b)=(195^\circ,-12^\circ)$ is the
    $\lambda$-Orionis ring). The dots show the brightest stars in the Orion
    constellation.  The circles show the positions of the three main subgroups
    of Ori~OB1.  From right to left are shown 1a, 1b and 1c.}
  \label{fig:orion}
\end{figure}
 
The Ori~OB1 association consists of four subgroups: 1a, located to the
northwest of Orion's Belt, 1b, located around the Belt, 1c, located around the
Sword of Orion, and 1d, which is the Trapezium cluster (see figure~7 in
\cite{Blaauw91}).  Brown et~al.\ \cite{Brown94} analyzed available Walraven
photometry \cite{Lub77} for Ori OB1. Model atmospheres and empirical
calibrations were used to determine effective temperatures, surface gravities,
and absolute bolometric magnitudes of the stars, and to derive ages for the
association subgroups by isochrone fitting in the Hertzsprung--Russell
diagram. It was found that the distances to subgroups 1a--1c are 380, 360, and
400~pc, respectively. The distance to 1d could not be determined reliably due
to the nebulosity in that region and insufficient stars. These distances were
smaller than the distances derived previously, a result which also followed
from a reanalysis of the data of Warren \& Hesser \cite{Warren77a} using a
revised calibration of the $uvby\beta$ system \cite{AT82}. The ages of the
Ori~OB1 subgroups were found to be $11.4\pm 1.9$~Myr for 1a, $1.7\pm 1.1$~Myr
for 1b, $4.6\pm 2$~Myr for 1c, and less than 1~Myr for subgroup 1d.  The IMF
was found to be a single power law with ${\rm d}\log(\xi(\log m))/{\rm d}\log
m = -1.7\pm0.2$ for all three subgroups (see also \cite{Brown98}).

Finally, we mention Cep~OB3, which is considered to be an example of
sequential star formation \cite{Elmegreen77}. The most recent studies of its
stellar content are by Jordi et~al.\ \cite{Jordi92}, \cite{Jordi96}.
Str\"omgren photometry was used to refine and extend the membership lists
towards fainter stars. The existence of two subgroups in this association was
confirmed and the ages were found to be $5.5$ and $7.5$~Myr.

\paragraph{Interstellar Medium.}
The nearby OB associations offer a uniquely detailed view of the relation
between early-type stars and the interstellar medium (cf.\ \S \ref{sec:why})
which helps us understand the impact of associations on the interstellar
matter throughout the Galaxy. Much work has been done since 1991,
concentrating especially on the interstellar medium around Ori~OB1. The
characteristics of the interstellar medium related to Sco~OB2 were summarized
in \cite{Blaauw91} and \cite{DeGeus92}.

For summaries of studies of the interstellar gas in the vicinity of Ori~OB1 we
refer to \cite{McCaughrean98}. The impact of the early-type stars in this
association has been addressed recently in a series of studies,
\cite{Brown95}, \cite{Burrows93}, \cite{Guo95}, and \cite{Snowden95}, taking
advantage of the availability of surveys such as IRAS, the ROSAT all sky
survey, and the Leiden--Dwingeloo \HI{} survey \cite{Hartmann97}.
Figure~\ref{fig:orion} shows the distribution of \HI{} around Orion over the
velocity range $-1$ to $+8$~km~s$^{-1}$. One can clearly distinguish a cavity
surrounded by a shell of \HI{}, the Orion--Eridanus bubble. The same features
can be seen in an image at 100~$\mu$m from IRAS \cite{Brown95}, and the whole
of the cavity is filled with very hot gas, $\sim$$10^6$~K, emitting in X-rays
\cite{Burrows93}, \cite{Brown95}, \cite{Snowden95}. The shell has a measured
expansion velocity of about $40$~km~s$^{-1}$ and a mass of $2.3\pm0.7\times
10^5$~M$_\odot$. Taking into account the IMF and the ages of the subgroups,
the mechanical energy output in the form of stellar winds and supernovae over
the lifetime of the association was estimated to be $\sim$$10^{52}$ ergs
\cite{Brown94}.  Using semi-analytic models of wind-blown bubbles that take
the density stratification of the Galactic \HI{} layer into account
\cite{Koo90}, it was shown that this energy is indeed enough to account for
the size as well as for the expansion velocity of the \HI{} shell
\cite{Brown95}.

Early-type stars also have a large effect on the interstellar medium through
their ionizing radiation, producing both localized \HII{} regions and diffuse
ionized gas. Based on the distribution of OB associations in the Galaxy, it
was shown that their luminosity function can be fit with a truncated power
law, and that there probably is a physical limit to the maximum size of \HII{}
regions in the Galaxy \cite{McKee97}.  A comparison with the distribution of
giant molecular clouds \cite{Williams97} showed that a 10$^6$~M$_\odot$ cloud
is expected to survive about $30$~Myr, and that on average 10 per~cent of its
mass is converted into stars by the time it is destroyed (see also the
chapters by Blitz and by McKee).  The overall distribution of associations is
also important for understanding the hot-gas filling factor of the
interstellar medium \cite{Ferriere95}--\cite{Ferriere98b}.

\paragraph{Formation of Stellar Clusters.}
Detailed studies of young clusters and OB associations lead to more insight
into the formation of these systems (cf.\ \S \ref{sec:why}). In an effort to
address this question the Orion Nebula Cluster (ONC), with the Trapezium
cluster at its core, was studied extensively by Hillenbrand
\cite{Hillenbrand97}. An optical sample of $\sim$$1600$ stars within $2.5$~pc
from the Trapezium stars was studied with photometry and spectroscopy. The
overall IMF of this core region of the ONC was found not to be grossly
inconsistent with `standard' stellar mass spectra. The observed IMF appears to
peak at $\sim$$0.2$~M$_\odot$ and to fall off rapidly towards lower masses.
Several substellar objects were identified. The total mass of the stars was
found to be $\sim$$1800$~M$_\odot$, and their mean age is less than 1~Myr,
with the younger stars being concentrated towards the centre. Mass segregation
was shown to be present in the ONC and it is probably not due to dynamical
relaxation of the cluster, implying that the massive stars formed near the
centre of the cluster. This confirms one of the predictions of cluster
formation theories as discussed in the chapter by Bonnell. A study of the
wider distribution of stars around the Trapezium shows that the ONC has an
elongated structure similar to that of the molecular gas distribution in the
region, suggesting that the cluster may still retain a memory of the geometry
of the protocluster cloud \cite{Hillenbrand98}.

\paragraph{Chemical Evolution.}
The interstellar medium in the Galaxy is continually enriched in chemical
elements by stellar winds and by ejecta from supernovae. OB associations are
generally still located near molecular clouds and thus are ideal sites for
studying ongoing chemical evolution processes in the Milky Way. Abundance
patterns of stars in Ori~OB1 have been studied by Cunha et~al.\ 
\cite{Cunha92}--\cite{Cunha98}. The abundance analysis shows that the stars in
Ori~OB1, in common with the Orion Nebula \HII{} region, are underabundant in
Oxygen with respect to the Sun. The lowest abundances are found in subgroups
1a and 1b.  The Trapezium stars and some stars of subgroup 1c seem to have O
abundances that are up to 40 per~cent higher than those in subgroups 1a and 1b
(although still subsolar). It is suggested that this is due to enrichment of
the interstellar gas by mixing of supernovae ejecta from subgroup 1c with the
gas that subsequently collapsed to form the Trapezium Cluster \cite{Cunha92}.
This enrichment scenario is confirmed by the fact that Cunha et~al.\ 
\cite{Cunha94} observe no abundance variations for C, N, and Fe, but do
observe the same variations for Si as for O, as one would predict for cloud
material enriched by Type~II supernova ejecta. Supernovae must have occurred
in Ori~OB1 in the past given the presence of the Orion--Eridanus bubble. It is
estimated that 1 to 2 supernovae have occurred in subgroup 1c \cite{Brown94}.

\paragraph{Kinematic Ages.}
Because OB associations are unbound they will expand \cite{Ambartsumian49}. In
principle one can use this expansion to trace back the motions of the stars in
an OB association until some minimum configuration is reached (see e.g.,
figure~5 in \cite{Blaauw91}). The time at which this happens would then
correspond to the kinematic age of the association, which can be compared to
the age derived from the Hertzsprung--Russell diagram. Also, an estimate is
obtained of the initial configuration of the association just after it was
formed. Kinematic ages have in fact been determined for a number of
associations \cite{Blaauw78}, \cite{Blaauw83}, \cite{Garmany73},
\cite{Lesh68}, and \cite{Lesh69}. However, it was demonstrated that tracing
back proper motions in OB associations always leads to underestimated ages and
overestimated initial sizes \cite{Brown97a}. The main reason is that the space
motions of the stars are not rectilinear but are influenced by the N-body
interactions in the initially more compact association, by the effects of the
remnant molecular cloud, and by the Galactic tidal field.

\paragraph{Binaries.}
Among the most important clues to understanding the process of star formation
are the characteristics of the binary population. Recently, a number of
searches for spectroscopic binaries in Sco~OB2 and Ori~OB1 were carried out
\cite{Levato87}, \cite{Morrell91}, and \cite{Verschueren96}, and the
statistics of close binaries among early-type stars were summarized by Blaauw
\cite{Blaauw91}. The problem in studying the binary population among
early-type stars in OB associations is to obtain precise radial velocities for
these stars, which is very difficult due to stellar rotation and the small
number of spectral lines \cite{Verschueren99}.  Nevertheless, we can expect
progress to be made in the near future when Hipparcos data on binaries is
analyzed.

With the advent of speckle and adaptive optics techniques, both in the optical
and infrared, much attention has been devoted recently to studying the
population of binaries among the stars in the Trapezium cluster and the
pre-main sequence stars in Sco~OB2 \cite{Petr98}, \cite{Brandner98}. Both
studies conclude that the characteristics of the binary population depend on
the star forming environment. In particular, Brandner \& K\"ohler
\cite{Brandner98} suggest that the conditions which favor the formation of
high-mass stars apparently lead to the formation of close binaries among
low-mass stars, whereas conditions unfavorable to the formation of high-mass
stars lead to the formation of wider binaries among low-mass stars. This bears
directly on the issues discussed in the chapter by Bonnell in this volume.

\subsection{Low-Mass Stars}\label{sec:lowmass}

Extrapolating the mass function for OB stars, such as derived in Sco~OB2 and
Ori~OB1 \cite{DeGeus92}, \cite{Brown94}, reveals that the bulk of the stars
should be of low mass ($\lesssim$$2$~M$_\odot$).  Indeed, evidence was found
early on for the presence of low-mass stars in the vicinity of the Orion
Nebula \cite{Haro53}, \cite{Walker69}. A search of the IRAS Point Source
Catalogue for young stellar objects \cite{Prusti92} shows how the distribution
of these stars clearly outlines some of the OB associations, such as Ori~OB1
and the Upper Scorpius (US) subgroup of Sco~OB2. Large-scale proper motion
searches for these fainter members of OB associations suffer from a much
larger contamination by field stars, making it hard to identify the members
without additional information. Hence, other techniques are employed in the
search for low-mass members of OB associations. The two most widely used are
objective prism H$\alpha$ surveys, which are sensitive to classical T-Tauri
stars, and X-ray surveys, also sensitive to weak-line T-Tauri stars (see the
chapter by M\'enard \& Bertout).

An EINSTEIN observatory X-ray search was used to look for the low-mass
population of US \cite{Walter94}. After correcting for the incompleteness of
the X-ray sampling, it was concluded that the association has a field star
mass function between about 0.2 and 10~M$_\odot$.  The total number of
low-mass stars ($< 2$~M$_\odot$) is about 2000, which is in good agreement
with the number of low-mass stars inferred by de Geus \cite{DeGeus92}. Recent
X-ray studies, based on the ROSAT all-sky survey \cite{Preibisch98},
\cite{Sciortino98} reveal the presence of many more X-ray selected pre-main
sequence (PMS) candidates throughout Sco~OB2.  Preibisch et~al.\ 
\cite{Preibisch98} performed spectroscopic follow-up observations to confirm
the PMS character of the X-ray selected sources and also looked for PMS stars
among objects without X-ray detections but with proper motions suggesting that
they might be members of Sco~OB2. No PMS stars were found among these proper
motion selected candidates and these authors conclude that the X-ray selected
sample of PMS stars is at least 75 per~cent complete.  However, Sciortino
et~al.\ \cite{Sciortino98}, using ROSAT HRI, PSPC and all-sky survey
observations, showed that the observations by Walter et~al.\ \cite{Walter94}
were much more incomplete than reported. They concluded that EINSTEIN and
ROSAT data are far from giving a complete characterization of the X-ray
population of US, implying the presence of even more PMS stars in Sco~OB2.

To confirm the PMS nature of the objects detected in X-ray surveys the
so-called Lithium test is used. Essentially one measures the strength of the
Li line which is correlated with stellar age for PMS stars. Main sequence
stars should have depleted all their Li in nuclear burning processes, and thus
the presence of Li indicates the PMS character of the object under study.
However, it has recently been pointed out that the Li test is not completely
reliable \cite{Favata96}--\cite{Favata98}.  Especially, when using
low-resolution spectra to study the Li line one may confuse young and active
main sequence stars with bona fide PMS stars, thus further complicating
studies of the IMF down to the lowest masses.

In Ori~OB1 the Kiso H$\alpha$ survey \cite{Nakano95a}, and the EINSTEIN
\cite{Walter97}, \cite{Walter98} and ROSAT \cite{Alcala96} X-ray surveys have
uncovered hundreds of emission-line and X-ray sources of which many are likely
to be PMS or T-Tauri stars. Spectroscopic follow-up observations indeed
confirmed the PMS nature of many of these stars \cite{Alcala96},
\cite{Kogure92}, \cite{Nakano95b}. However, it was shown \cite{Alcala98} that
the population of X-ray sources towards Orion consists of a mixture of true
Orion PMS stars and young foreground stars. The latter may be related to
Gould's Belt or may be $\sim$$10^8$~yr old stars.

Recent studies have revealed a significant population of low-mass PMS stars in
the region of subgroup 1b of Ori~OB1 \cite{Walter97}, \cite{Walter98}. The
stars appear to cluster spatially around $\sigma$~Ori and the narrowness of
their PMS locus suggests coevality, at the 2~Myr age of Ori~OB1b.  The total
inferred mass of this group of stars is comparable to that of the Orion Nebula
Cluster. In fact the $\sigma$~Orionis cluster may be an older analogue of the
Trapezium cluster. This result illustrates the difficulty in defining subgroup
boundaries in associations (cf.\ \S \ref{sec:define}). It may be that subgroup
1b actually consists of a `merger' of several Trapezium-like clusters. Most
recently, the detection of brown dwarf candidates around $\sigma$~Ori has been
reported. We refer to \cite{Walter99} for the details and more material on
low-mass stars in OB associations.

Enormous progress has been made towards characterizing the low-mass population
in OB associations. The challenge now is to bring the studies of the high- and
low-mass stars together into a comprehensive picture.

\section{Results from Hipparcos}\label{sec:census}

We now review the results on the stellar content of the nearby OB associations
derived from analysis of Hipparcos data. The material in this section is
essentially a summary of \cite{Z99}, \cite{DeBruijne99}, and
\cite{Hoogerwerf99} and we refer to those papers for all the details and
further references.

The most important data contained in the Hipparcos Catalogue \cite{ESA97} are
positions, proper motions, and trigonometric parallaxes for $\sim$120$\,$000
stars. The median precision of the astrometric data for stars brighter than
$V$$\sim$9 is $0.88$ and $0.74$~milli-arcsecond yr$^{-1}$ for the proper
motions in right ascension and declination, respectively, and
$0.97$~milli-arcsecond (mas) for the parallaxes. Note that at the distance of
the nearest associations ($\sim$$150$~pc) these numbers translate into
individual stellar distances accurate to 15 per~cent and transverse velocities
accurate to $0.6$~km~s$^{-1}$. The astrometric data are supplemented with
magnitudes and colors, variability information obtained from the photometry,
and detailed data on the properties of binaries observed or discovered by
Hipparcos.  Simultaneously with the main mission the Tycho experiment was
carried out.  This resulted in a catalogue of over 1 million stars to
$V\approx11.5$ containing astrometric data with less precision (median
$\sim$7~mas for $V\lesssim9$) as well as accurate photometry.

The Hipparcos Catalogue has a limiting magnitude of $V$$\sim$12, and is
complete to $V$$\sim$7.3 in the Galactic plane, and to $V$$\sim$9 in the polar
regions.  Furthermore, for reasons having to do with the way the observations
were carried out, the Catalogue suffers from severe selection effects in some
regions of the sky. For OB associations the relevant selection biases are
described in \S 3 of de Zeeuw et~al.\ \cite{Z99}. For more details on the
Hipparcos Catalogue please refer to the extensive documentation provided with
the Catalogue itself \cite{ESA97}.

\subsection{Astrometric Membership Selection}\label{sec:members}

Two methods, based on the assumption of common space motions for stars, were
used to select astrometric members of the nearby OB associations \cite{Z99}.
The first is a modification of the classical convergent point method
\cite{DeBruijne99}, and the second is a new selection method which makes use
of the parallaxes as well as the proper motions, and searches for members in
velocity space \cite{Hoogerwerf99}. Both methods were combined to find the
members of the nearby associations. This leads to a very powerful membership
selection tool in which many spurious interlopers that occur when either
method is used alone are automatically removed. Even so, one should still
expect that a fraction of the selected stars actually belong to the field but
by coincidence have the right motion to be included as members. The fraction
of expected interlopers was estimated for each association through extensive
Monte Carlo simulations.

\begin{figure}[htb]
  \begin{center} 
    \epsfig{file=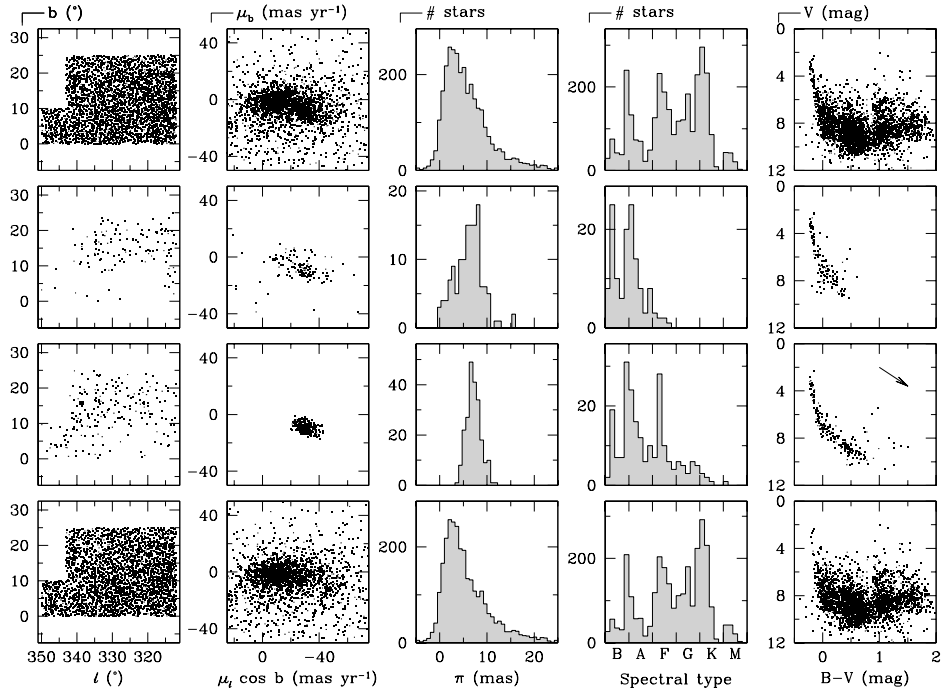,width=\hsize}
  \end{center}
  \caption{Hipparcos measurements for the subgroup Upper Centaurus
    Lupus of Sco OB2 (from the top row down): (1) all 3132 Hipparcos stars in
    the region; (2) the 136 pre-Hipparcos members; (3) the 221 Hipparcos
    members; (4) the remaining stars after member selection. The columns show
    (from left to right): (1) positions in Galactic coordinates; (2) Galactic
    vector point diagram; (3) trigonometric parallax distribution; (4)
    spectral type distribution; (5) color-magnitude diagram, not corrected for
    reddening.  The arrow indicates the direction of reddening for the
    standard value $R=3.2$ of the ratio of total to selective extinction.}
  \label{fig:ucl}
\end{figure}

To illustrate the dramatic advances provided by the Hipparcos data,
Figure~\ref{fig:ucl} shows in detail the membership selection results for the
Upper Centaurus Lupus (UCL) subgroup of Sco~OB2. The first row displays the
Hipparcos measurements for all stars in the UCL field.  The panels show no
clear sign of a physical group, except for the vector point diagram (panel
two), which contains a concentration around $(\mu_{\ell} \cos b,
\mu_{b})\sim(-25, -10)$~mas~yr$^{-1}$ superimposed upon the broader Galactic
disk distribution.  The second row of Figure~3 shows 136 stars that were
proposed as members of UCL, based on pre-Hipparcos kinematic and photometric
studies. They are mostly B- and A-type stars, with fairly little concentration
in the vector point diagram. Their parallax distribution is narrower than the
one in the first row, and peaks around 9~mas. The characteristics of the set
of astrometrically selected members are presented in the third row of
Figure~\ref{fig:ucl}. The vector point diagram of these secure members is much
more concentrated than that of the pre-Hipparcos members, and the parallax
distribution is narrower. This is due to a reduced contamination by field
stars. The spread and the elongated shape in the vector point diagram are
consistent with the combined effects of observational errors, the estimated
internal velocity dispersion, and projection on the sky \cite{Z99}.  Note how
much more extended the HR-diagram, not corrected for reddening, of the
association is after Hipparcos membership selection. The data now extend to
beyond spectral type F and in fact some PMS objects are included (see \S
\ref{sec:nearbyfut}). Finally, the panels in the bottom row of
Figure~\ref{fig:ucl} show the not-selected stars, and demonstrate that the
membership selection procedure does not leave `holes' in the distributions of
positions, proper motions, and parallaxes. This indicates that UCL was
separated cleanly from the field stars.

\begin{figure}[thb]
  \begin{center}
    \epsfig{file=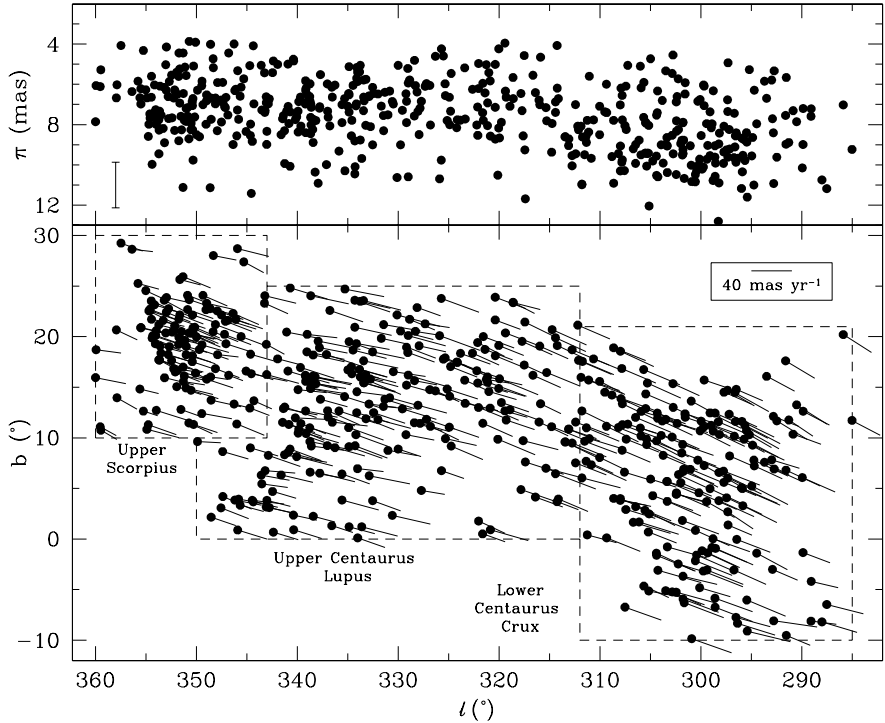,height=9.75truecm}
  \end{center} 
  \caption{Positions and proper motions (bottom), and
    parallaxes (top), for 521 members of Sco~OB2 selected from 7974 stars in
    the Hipparcos Catalogue in the area bounded by the dashed lines. The
    vertical bar in the top panel corresponds to the average $\pm1\sigma$
    parallax range for the stars shown. US is identified as a subgroup based
    on the concentration of members on the sky, and LCC can be distinguished
    from US and UCL based on its significantly smaller distance.}
  \label{fig:scoob2}
\end{figure}

\subsection{Selected Results}\label{sec:selected}

The outcome of the Hipparcos membership selection is described below with a
couple of specifically chosen examples from de Zeeuw et~al.\ \cite{Z99}, meant
to illustrate the various ways in which high-precision proper motion surveys
can impact the field of OB associations. The case of Sco~OB2 was chosen to
show that the membership lists can now be extended into the PMS regime. The
findings on Vel~OB2 and Tr~10 show the dramatic improvements in the ability to
identify OB associations (Vela), and how in some cases what was thought to be
an open cluster actually turns out to be an association (Trumpler~10). The
results on Col~121 show that we can now detect analogues of Sco~OB2 at much
larger distances (a factor four in this case). Cep~OB6 is an example of a
newly discovered association. Finally, we discuss Ori~OB1 as an example of
some of the difficulties encountered in identifying the members of
associations.

\paragraph{Scorpius~OB2.}
Figure~\ref{fig:scoob2} illustrates the proper motions of all the Sco~OB2
members that have been identified, and also gives the subgroup boundaries. In
US a total of 120 members were found located in a volume of $\sim$30~pc
diameter at a mean distance of $145\pm2$~pc (see \S \ref{sec:distances} for
more information on the determination of mean distance). In UCL 221 members at
$140\pm2$~pc were found, and in Lower Centaurus Crux (LCC) 180 at
$118\pm2$~pc.

Sco~OB2 clearly forms one coherent structure, although US stands out in the
distribution of Sco~OB2 members on the sky, and the parallax distribution
clearly distinguishes UCL and LCC\@. The differences between the
Hertzsprung--Russell diagrams of the groups \cite{DeGeus89} also indicate that
a division of Sco~OB2 into three separate subgroups is warranted. The field
boundary separating UCL from US has, somewhat arbitrarily, been chosen in such
a way that US comprises the stellar concentration centered on
$(\ell,b)\sim(352^\circ\!,20^\circ\!)$ with radius $\sim5^\circ\!$.

\begin{figure}[thb]
  \begin{center}
    \epsfig{file=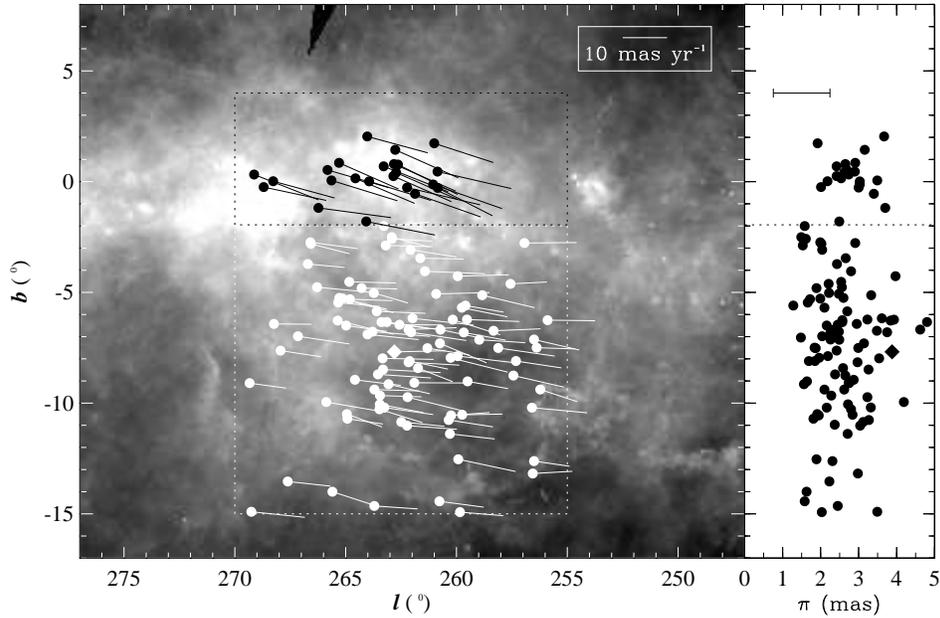,width=\textwidth}
  \end{center}
  \caption{Left: positions and proper motions for the Hipparcos members of
    Vela~OB2 (white) and Trumpler~10 (black). The diamond denotes the
    Wolf--Rayet star $\gamma^2$~Velorum (WR11). The dotted lines indicate the
    field boundaries. The grey scale represents the IRAS 100~$\mu$m skyflux.
    The IRAS Vela shell is the ring-like structure centered on $(\ell,
    b)\sim(263^\circ\!, -7^\circ\!)$ with a radius of $\sim$6$^\circ\!$
    surrounding Vel~OB2. The intense emission in the area $260^\circ\!\!
    \lesssim \! \ell \! \lesssim \! 273^\circ\!$, $-2^\circ\!\! \lesssim \! b
    \! \lesssim \! 2^\circ\!$ corresponds to the Vela molecular ridge. Right:
    parallax distribution for Vel~OB2 and Tr~10.}
  \label{fig:velaob2}
\end{figure}

\paragraph{Vela OB2 and Trumpler 10.}
In his 1914 paper, Kapteyn not only identified Sco~OB2, but also discussed a
group of bright stars in Vela and, based on proper motions, listed 15 probable
members for this so-called Vela Group \cite{Kapteyn14}. Ever since a clear
identification of the members of this group has remained difficult \cite{Z99}.
However, the new Hipparcos results, illustrated in Figure~\ref{fig:velaob2},
show very clearly the presence of an OB association in the Vela region. In
fact 93 members have been identified of which 89 are new! The color-magnitude
diagram of this association shows that the earliest spectral type on the main
sequence is B1, suggesting an age of $\lesssim$10~Myr. The association is
located at a distance of $410\pm12$~pc, and the new members are concentrated
on the sky around $(\ell, b)\sim(263^\circ\!, -7^\circ\!)$ within a radius of
$\sim$5$^\circ\!$. Sahu \cite{Sahu92} reported the detection of the so-called
IRAS Vela shell in the IRAS Sky Survey Atlas maps (cf.\ \cite{Sahu94}). This
is an expanding shell, centered on Vel~OB2, with a projected radius of
$\sim$6$^\circ\!$ (Figure~\ref{fig:velaob2}).  Sahu showed that the observed
kinetic energy of the IRAS Vela shell is of the same order of magnitude as the
total amount of energy injected into the interstellar medium through the
combined effects of stellar winds and supernovae if the shell were to contain
a `standard' OB association \cite{Sahu92}.  The subsequent astrometric
identification of Vel~OB2 as a rich OB association, and the improved distance
determination, confirm the relation of the association with the Vela shell.

Based on relative proper motion data for 29 stars, Lyng\aa\ \cite{Lynga59},
\cite{Lynga62} identified 19 probable members of the sparse open cluster
Trumpler~10. Analysis of the Hipparcos data showed that Tr~10 is actually a
moving group. The Catalogue contains 23 members: 22 B-type stars (earliest
spectral type B3V) and 1 A0V star. Figure~\ref{fig:velaob2} shows the members
of Tr~10 seen in projection in front of the Vela molecular ridge. Tr~10 is
located at a distance of $366\pm23$~pc. Based on its color-magnitude diagram
this group is clearly older than Vel~OB2, and a provisional age estimate based
on the earliest spectral type is $\sim$15~Myr.

\begin{figure}[thb]
  \begin{center}
    \epsfig{file=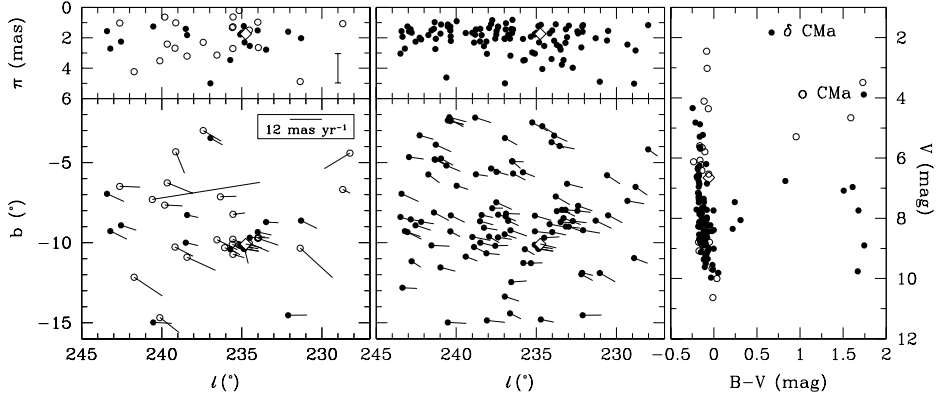,width=\hsize}
  \end{center}
  \caption{Left: positions and proper motions (bottom) and parallaxes (top) of
    the pre-Hipparcos members of Collinder~121. Filled circles are confirmed
    members. Open circles indicate stars not considered members after analysis
    of Hipparcos data. The confirmed member EZ~CMa (WR6, see
    \protect\cite{Z99}) is indicated by an open diamond. Middle: same diagram
    for all stars selected as member of Col~121, illustrating the dramatic
    change from pre- to post-Hipparcos. Right: color-magnitude diagram, not
    corrected for reddening, for the Col~121 members (filled circles), and
    rejected classical members (open circles).  The unusual position of EZ~CMa
    (open diamond) may be caused by systematic effects in the Hipparcos
    photometry of this star \protect\cite{Z99}.}
  \label{fig:col121}
\end{figure}

\paragraph{Collinder 121.}
Collinder \cite{Collinder31} studied the structural properties and spatial
distribution of Galactic open clusters, and discovered a cluster of 20 stars
at $\sim$590~pc in an area of $1^\circ\! \times 1^\circ\!$ on the sky:
Col~121. Schmidt--Kaler \cite{SK61} noted a large number of evolved early-type
stars in a field of $10^\circ\!  \times 10^\circ\!$ centered on the bright
supergiant $o$~CMa located in the central part of Col~121. Subsequent studies
revealed the possible existence of an OB association in this field \cite{Z99}.

This is indeed confirmed by the Hipparcos results. 103 stars were selected in
the Col~121 field with a mean distance of 592$\pm$28~pc.
Figure~\ref{fig:col121} shows both the stars previously considered members of
this group as well as the newly selected Hipparcos members. The figure also
presents the Hipparcos color-magnitude diagram, not corrected for reddening.
It shows that Col~121 contains a number of evolved stars. The abrupt cutoff of
the main sequence near $V$=10 is caused by the completeness limit of the
Hipparcos Catalogue. Some of the late-type stars may well be interlopers.  The
presence of an O star and early-type B stars indicates this is a young group,
of age $\sim$5~Myr.

Col~121 has completely changed its appearance compared to the classical
membership lists. The middle panel of Figure~\ref{fig:col121} suggests two
subgroups, $(\ell,b)$ $\sim$ $(233^\circ\!,-9^\circ\!)$ and
$(238^\circ\!,-9^\circ\!)$. A possible third subgroup lies at $(\ell,b)$
$\sim$ $(243^\circ\!,-9^\circ\!)$.  The linear dimensions of this complex are
100$\times$30 pc, similar to that of the Sco OB2 association.  The
color-magnitude diagram also resembles that of Sco OB2, in the sense that the
earliest spectral type and amount of extinction are similar.

\begin{figure}[h]
  \begin{center}
    \epsfig{file=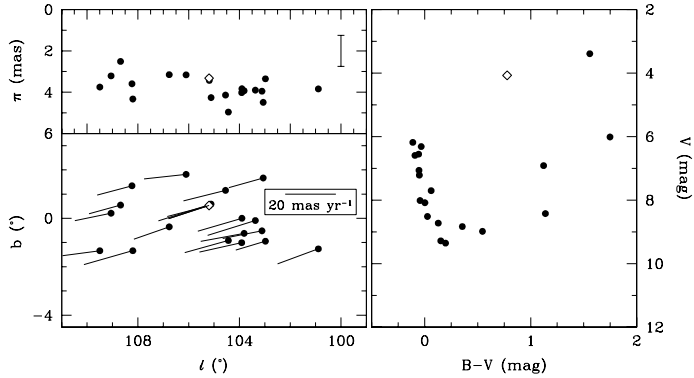,width=9.2truecm}
  \end{center}
  \caption{Left: positions and proper motions (bottom) and parallaxes (top)
    for the new moving group in Cepheus, designated Cepheus~OB6. Right:
    color-magnitude diagram, not corrected for reddening. The open diamond
    denotes $\delta$~Cep.}
  \label{fig:cepob6}
\end{figure}

\paragraph{Cepheus OB6.}
Following a discovery by Hoogerwerf et~al.\ \cite{Hoogerwerf97}, de Zeeuw
et~al.\ \cite{Z99} examined the field $100^\circ\!\! < \! \ell \! < \!
110^\circ\!$ and $-4^\circ\!\!  < \! b \! < \! 3^\circ\!$ in the Cepheus
region and found a previously unknown moving group. This group, designated as
Cepheus~OB6, consists of 20 members. Their positions, proper motions and
parallaxes are shown in Figure~\ref{fig:cepob6}. The color-magnitude diagram
is very narrow and strengthens the evidence that these stars form a moving
group. It is probably an old OB association: the earliest spectral type is
B5III, suggesting an age of $\sim$50~Myr. A noteworthy member of this group is
$\delta$~Cep, the prototype classical Cepheid.

\begin{figure}[h]
  \begin{center}
    \epsfig{file=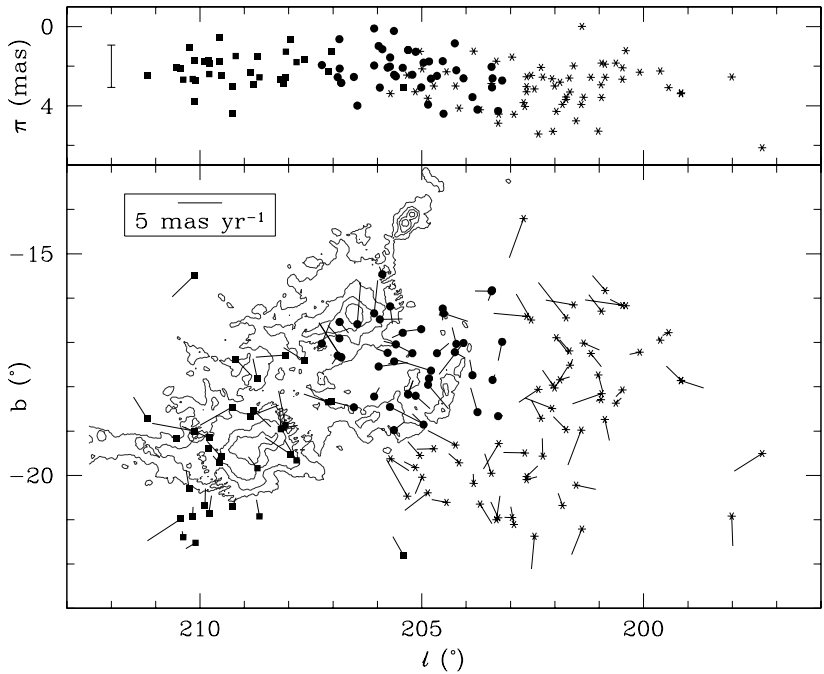,height=9.0truecm}
  \end{center}
  \caption{Positions and proper motions (bottom) and
    parallaxes (top) for the stars of Ori~OB1 selected by Brown et~al.\ 
    \protect\cite{Brown98}. The proper motions are small because Ori~OB1 lies
    near the direction of the Solar Antapex. The parallaxes in subgroup 1a
    (asterisks) are generally larger than those in 1b (filled circles) and 1c
    (filled squares). The contours indicate the 100~$\mu$m IRAS flux map.}
  \label{fig:oriob1}
\end{figure}

\paragraph{Other Associations.}
Astrometric membership detection for OB associations with Hipparcos data was
successful out to $\sim650$~pc. The following associations were also detected:
Per~OB2, $\alpha$~Persei (Per~OB3), Cas--Tau, Lac~OB1, and Cep~OB2.
Astrometric evidence for moving groups in the fields of R~CrA, CMa~OB1, Mon
OB1, Ori~OB1, Cam~OB1, Cep~OB3, Cep~OB4, Cyg~OB4, Cyg~OB7, and Sct~OB2, is
inconclusive. OB associations do exist in many of these regions, but they are
either at distances where the Hipparcos parallaxes are of limited use, or they
have unfavorable kinematics, so that the group proper motion does not
distinguish them from the field stars in the Galactic disk.

Among the latter group we note especially Ori~OB1. It is the best-studied
nearby OB association but unfortunately it is located close to the Solar
Antapex, and its space motion is almost purely radial with respect to the Sun.
This means that it is very difficult to select members in Ori~OB1 based on
proper motions. A discussion on Ori~OB1 is given in \cite{Z99} based on the
preliminary study described in detail in \cite{Brown98}. In that study a very
crude selection of Ori~OB1 members was done, based on Hipparcos proper
motions, and the mean distances to the subgroups were calculated. These
distances are: 336$\pm$16~pc for 1a, 473$\pm$33~pc for 1b and 506$\pm$37~pc
for 1c, where the quoted errors are the formal errors on the mean distances.
The actual uncertainty is larger due to the simplified member selection. All
of this is illustrated in Figure~\ref{fig:oriob1} which shows the proper
motions and parallaxes of the selected members. Note that from the parallaxes
alone it is clear that subgroup 1a is located much closer to the Sun than 1b
and 1c \cite{Brown98}.

\subsection{Mean Distances and Motions}\label{sec:distances}

Use of the Hipparcos trigonometric parallaxes of the secure members to
determine mean distances to the associations or their subgroups requires some
caution, as the inverse of the parallax is a biased distance indicator
\cite{Smith96}, \cite{Brown97b}, and the conversion of mean parallax to mean
distance for a group of stars depends on the distribution of stars within the
group. It can be shown \cite{Z99} that for all spherical groups the
expectation value of the mean of the measured parallaxes is equal to the true
mean parallax, and corresponds to the true distance of the centre of the
group. For elongated associations the bias in the mean parallax is small,
typically less than 1 per~cent. Hipparcos parallaxes measured in regions of
high stellar density (in the Catalogue) have to be interpreted with care
\cite{Lindegren89}, \cite{Pinsonneault98}, \cite{Robichon97}. This is not a
problem for the low-density associations.

Furthermore, the observed distribution of parallaxes may not be representative
of the true underlying parallax distribution of an association. Magnitude
limits and the characteristics of the Hipparcos Input Catalogue bias the
selected members (see \cite{Z99} for more details), and some stars in the
Hipparcos Catalogue have a negative measured parallax. The member selection
method described in \S \ref{sec:members} rejects these, which introduces a
bias towards a smaller mean distance. De Zeeuw et~al.\ \cite{Z99} were able to
correct the mean distances to the associations for this bias, by carrying out
Monte Carlo simulations to estimate its magnitude.

The resulting distances and projected sizes of the kinematically detected
nearby OB associations are listed in Table~\ref{tab:distance}. This table also
lists the mean proper motions. The means are based on the individual
measurements for the association members. The Hipparcos Input Catalogue lists
radial velocities from a variety of sources. They are not available for all
secure association members, and therefore Table~\ref{tab:distance} lists the
median radial velocity, and no error estimates are given.

The mean distances derived in Table~\ref{tab:distance} have very small formal
errors due to the averaging over large numbers of stars. However, associations
have large physical sizes, and the mean distances listed here should not be
taken as `the distance' to all association members, but rather as an
indication of the location of their centroid.

The comparison between the new Hipparcos distances and the distances in the
literature for OB associations reveals a tight correlation, although the
Hipparcos distances are systematically smaller by about $0.2$ magnitudes in
the distance modulus. This issue is discussed in more detail by de Zeeuw
et~al.\ \cite{Z99}.

To conclude we return to the issues discussed in \S \ref{sec:define}.  The
projected sizes of the associations correlate roughly with their distances.
This is obviously not a real physical effect but merely a reflection of how
the field boundaries were established prior to membership selection. The case
of Col~121 is a good illustration of how an association like Sco~OB2, which
consists of three subgroups, may resemble a single entity at larger distances
(see Figure~\ref{fig:col121}). Moreover, any kinematic distinction between
subgroups within an association is difficult to establish (cf.\ 
Figure~\ref{fig:scoob2}), and extra information, such as photometry, is
required. Thus, even though OB associations may possibly form with particular
characteristic dimensions one should always bear in mind the observational
biases that enter when addressing the question of initial configurations of
young star clusters.

\begin{table}[htb]
\begin{center}
\caption{\label{tab:distance}
Mean distances and mean motions of the nearby OB associations}
\begin{tabular}{lrrrrrrrr}
\hline
Name & $D$ & $N$ & Size & $\langle \mu_\ell \cos b\rangle$ &
$\langle \mu_b\rangle$ & $v_{\rm rad}$ & Age & $N_{\rm br}$ \\
\hline
US &           $145\pm\phn2$ & $120$ &  $30$ & $-24.5$ & $-8.1$  &  $-4.6$ & $5$      & $5$ \\
UCL &          $140\pm\phn2$ & $221$ &  $65$ & $-30.1$ & $-9.1$  &   $4.9$ & $13$     & $10$ \\
LCC &          $118\pm\phn2$ & $180$ &  $45$ & $-32.1$ & $-13.1$ &  $12.0$ & $10$     & $4$ \\
Vel~OB2 &      $410\pm12$    &  $93$ &  $70$ & $-10.4$ & $-1.3$  &  $18.0$ & $10$     & $12$ \\
Tr~10 &        $366\pm23$    &  $23$ &  $45$ & $-14.3$ & $-4.9$  &  $21.0$ & $15$     & $0$ \\
Col~121 &      $592\pm28$    & $103$ & $115$ & $-5.1$  & $-1.5$  &  $26.0$ & $5$      & $24$ \\
Per~OB2 &      $318\pm27$    &  $41$ &  $45$ &  $8.4$  & $-2.3$  &  $20.1$ & $4$--$8$ & $2$ \\
$\alpha$~Persei & $177\pm\phn4$ &  $79$ &  $15$ & $33.5$  & $-8.7$  &  $-1.0$ & $50$     & $2$ \\
Lac~OB1 &      $368\pm17$    &  $96$ &  $65$ & $-2.3$  & $-3.4$  & $-13.3$ & $16$     & $7$ \\
Cep~OB2 &      $615\pm35$    &  $71$ & $105$ & $-4.1$  & $-0.5$  & $-21.4$ & $5$      & $11$ \\
Cep~OB6 &      $270\pm12$    &  $20$ &  $40$ & $15.9$  & $-4.4$  & $-20.0$ & $50$     & $2$ \\
\noalign{\vskip5pt}
Ori~OB1a$^\star$ & $336\pm16$ & $61$ & $30$ & $0.8$ & $0.1$ & $-23.0$ & $11$ & $6$ \\
Ori~OB1b$^\star$ & $473\pm33$ & $42$ & $30$ & $0.8$ & $0.1$ & $-23.0$ & $\phn2$ & $9$ \\
Ori~OB1c$^\star$ & $506\pm37$ & $34$ & $30$ & $0.8$ & $0.1$ & $-23.0$ & $\phn5$ & $12$ \\
\hline
\end{tabular}
\end{center}

All distances, in pc, include a correction for systematic effects as described
in \S \ref{sec:distances}. The first column lists the association. The second
column contains the mean distance and the error on the mean. The third column
lists the number of Hipparcos members. Column 4 lists the projected size in pc
of the associations derived from their extent on the sky and assuming
spherical symmetry. Columns~5 and 6 list the average proper motions in the
directions of Galactic longitude and latitude. The formal errors on the mean
are typically $0.1$--$0.2$~${\rm mas~yr^{-1}}$.  Column~7 contains the median
radial velocity in ${\rm km}~{\rm s^{-1}}$ compiled from the Hipparcos Input
Catalogue. Column~8 contains the (approximate) age in Myr; see \cite{Z99} for
references. Column~9 contains the number of stars with absolute magnitude
brighter than $-2$. The absolute magnitudes were calculated from the mean
distances and values for $V$ as listed in the Hipparcos Catalogue. Due to its
large physical size and unknown orientation, the Cas--Tau association cannot
be represented adequately by a single mean distance, proper motion, and radial
velocity, and has been excluded from this table.

$^\star$ The distances and numbers of members for Ori~OB1 are based on an {\it
  ad hoc\/} selection procedure and should be regarded as {\it provisional\/}
numbers only.

\end{table}

\section{Gould's Belt and the Origin of the Nearby Associations}
\label{sec:gould}

\begin{figure}[htb]
  \begin{center}
    \epsfig{file=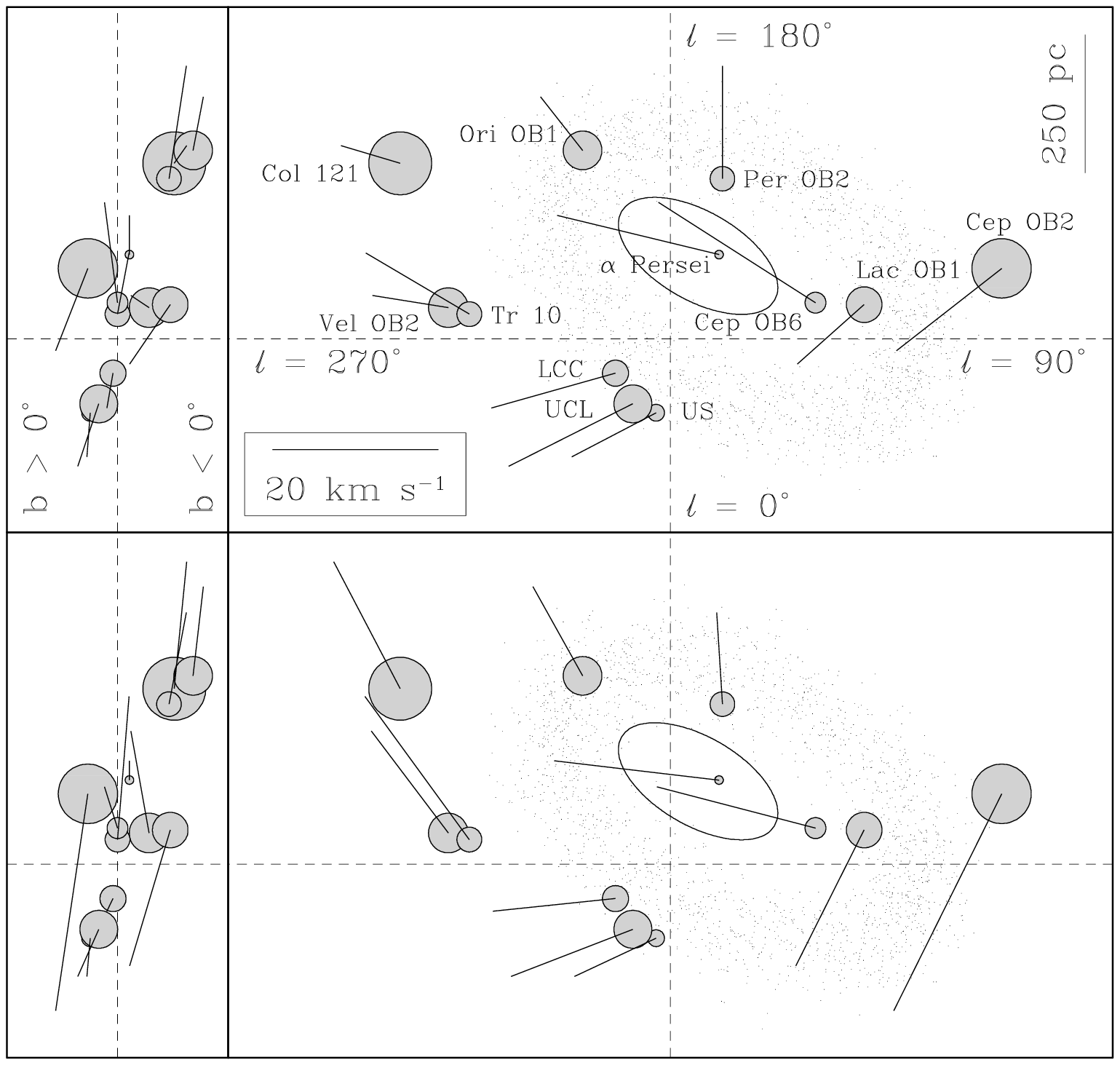,height=8cm}
  \end{center}
  \caption{Locations of the kinematically detected OB associations
    projected onto the Galactic plane (right, cf.\ Figure~\ref{fig:ruprecht})
    and a corresponding cross section (left). The grey circles indicate the
    physical dimensions as obtained from the angular dimensions and mean
    distances, on the same scale. The lines represent the streaming motions,
    derived from the average proper motions, mean distances and median radial
    velocities of the secure members, corrected for `standard' Solar motion
    and Galactic rotation (upper panel) or for Solar motion only (lower
    panel). The ellipse around the $\alpha$~Persei cluster indicates the
    Cas--Tau association. The small dots schematically represent a model of
    the Gould Belt \protect\cite{Olano82}.}
  \label{fig:gould}
\end{figure}

Mean space motions of the nearby OB associations, in km~s$^{-1}$, were derived
from the mean proper motions, mean distances, and the median radial velocities
\cite{Z99}. Figure~\ref{fig:gould} shows the result, after subtraction of
Solar motion \cite{Dehnen98} (lower panel) and, additionally, differential
Galactic rotation \cite{Feast97} (upper panel). When considering the motions
with respect to the Local Standard of Rest, some of the associations in
Figure~\ref{fig:gould} (lower panel) seem to fit a coherent pattern of
expansion and rotation, which is very similar to that derived from Hipparcos
measurements of OB stars with ages less than $\sim$30~Myr \cite{Lindblad97},
\cite{Torra97}. This large-scale feature is known as Gould's Belt, which is
the flat system of early-type stars within $\sim$500~pc \cite{Gould1874},
associated with a large structure of interstellar matter, including reflection
nebulae, dark clouds, and \HI{}. Its most striking feature is a tilt of
$\sim$18$^\circ\!$ with respect to the Galactic plane. It has also been
detected in the distribution of young stars observed in X-rays by ROSAT
\cite{Guillot98}. P\"oppel \cite{Poppel97} recently wrote a comprehensive
review of this structure, with emphasis on the role and characteristics of the
interstellar medium.

Following a suggestion by Blaauw \cite{Blaauw65}, who studied the mean space
motions of the nearby OB associations, Lindblad \cite{Lindblad67} interpreted
the observations of the local \HI{} gas associated with Gould's Belt in terms
of a ballistically expanding ring. Subsequent kinematic models combined with
new observations confirmed this picture (e.g., \cite{Elmegreen82},
\cite{Olano82}). However, these models seem inadequate for a description of
the space motions of the clumpy distribution of early-type stars, as no unique
expansion center and/or expansion age can be defined for them (e.g.,
\cite{Blaauw65}, \cite{Comeron94}, \cite{Lesh68b}, \cite{Westin85}).  The lack
of homogeneous and accurate radial velocities for all local early-type stars
prevents an optimal exploitation of the Hipparcos data, and a full kinematic
model is not yet available. Even so, we can use the mean motions of the OB
associations to shed light on the systemic motions in Gould's Belt.

The associations Sco~OB2, Ori~OB1, Per~OB2, and possibly Lac~OB1, are thought
to be components of the Gould Belt (e.g., \cite{Olano82}). The pattern
displayed in the lower panel of Figure~\ref{fig:gould} seems to be shared by
Tr~10, and perhaps also by Vel~OB2 and Col~121. Moreover, all these
associations are younger than $\sim20$~Myr, suggesting that these may all
belong to the same coherent structure. If it is part of Gould's Belt, the new
distance of Lac~OB1 reduces the extent of the Belt in the direction $\ell\sim
90^\circ\!$. Cep~OB2 is not located in the plane of the Belt, and its motion
is parallel to the Galactic plane. It does not seem to belong to the Belt. The
Cas--Tau complex which surrounds the $\alpha$~Persei cluster and shares its
motion, as well as Cep~OB6, are located inside the main ring of associations,
have a different space motion, and are all significantly older at
$\sim$50~Myr.

The following scenario was suggested for the formation of Gould's Belt and its
constituent associations \cite{Elmegreen93}, \cite{Elmegreen98}. The Carina
spiral arm is presently $\sim$$4$~kpc from the Sun along the Solar circle at
$\ell=282^\circ$ \cite{Graham70}. If the pattern speed of the associated
spiral density wave is $13.5$~km~s$^{-1}$ kpc$^{-1}$ \cite{Yuan69}, then the
physical speed of the local arm is 114~km~s$^{-1}$ in the tangential
direction. This means that the time since the passage of the Carina spiral arm
near the Sun would have been 35~Myr without streaming motions parallel to the
arm. With streaming, the time may be twice this value. As the oldest
components of Gould's Belt have a similar age, this suggests that the Belt
began as a giant gaseous condensation in the Carina spiral arm when the
location of the arm and the Sun last coincided. In this gas probably the
Cas--Tau, $\alpha$~Persei, Cep OB6, and Pleiades clusters formed, as well as
many dispersed B and later-type stars \cite{Lesh68b}, \cite{Poppel97}. The
combined action of the stellar winds and supernovae from the massive stars in
these older stellar groups may have led to the ring-like gaseous structure
from which the younger associations formed \cite{Blaauw91}. An alternative
scenario for the origin of Gould's Belt, invoking the oblique impact of a
large high-velocity cloud onto the Galactic disk, was put forward to explain
the tilt in Gould's Belt \cite{Comeron94}.

\section{Future Perspectives}\label{sec:future}

In his 1991 chapter on associations Blaauw concluded with the promise of major
advances in the field through the availability of Hipparcos data as well as
much improved radial velocities. Although precise radial velocities for the
early-type stars are not yet available, the Hipparcos data have now been
analyzed and have put the issue of association membership on much firmer
ground. Below we will discuss how this provides an excellent start for future
investigations and we briefly discuss what is in store for work on OB
associations beyond the Galaxy.

\subsection{The Nearby OB Associations}\label{sec:nearbyfut}

Hipparcos has provided a much improved description of the ensemble of young
stellar groups in the Solar neighbourhood, out to a distance of $\sim$650~pc.
The list of members in the astrometrically detected associations has in all
cases been refined and extended and a much better kinematical description of
Gould's Belt is now available. With the number of candidate members greatly
reduced, we are now in an excellent position to start gathering additional
data such as photometry and radial velocities, which will help in further
refining the membership lists and studying the physical properties of the
association members.

The discovery of Cep~OB6 in the Cep~OB2 field \cite{Z99} suggests that there
might be other previously unidentified nearby associations. A systematic
search in regions of the strip $-30^\circ\!\! \le \! b \!  \le \! 30^\circ\!$
not covered by de Zeeuw et~al.\ \cite{Z99} might reveal such groups. It will
also be interesting to investigate whether, as in the Cas--Tau/$\alpha$~Persei
pair, more open clusters exist with an extended halo, which appears as an old
association.

De Zeeuw et~al.\ \cite{Z99} presented color-magnitude diagrams based on the
$V$ and $B\!-\!V$ data in the Hipparcos Catalogue, but the physical parameters
of the member stars were not discussed in detail, mostly because the required
homogeneous multi-color photometry (and spectral classification) is
incomplete. It is now relatively easy to acquire the necessary photometry.
This is particularly interesting in Sco~OB2, where the faintest astrometric
members can be connected directly to the populations of PMS objects discovered
through X-ray searches. This will provide accurate ages, initial mass
functions \cite{Brown98b}, and the energy and momentum input into the
interstellar medium, thus enabling a detailed investigation of the influence
of young stellar groups on the surrounding distribution of gas and dust. The
Leiden--Dwingeloo \HI{} survey \cite{Hartmann97}, which has recently been
completed for the southern sky, can be used to undertake a comprehensive study
of the interstellar medium surrounding the nearby associations. It will also
be interesting for the nearest associations to combine the Hipparcos parallax
measurements with absorption-line measurements of gas in order to gain insight
into the 3D-structure of the interstellar medium.

An important next step is to complement the uniform astrometric measurements
provided by Hipparcos with homogeneous radial velocities with accuracies of
$\sim 3$~km~s$^{-1}$ or better, which is a considerable challenge for earlier
spectral types (cf.\ \S \ref{sec:highmass}). An unbiased member selection
based only on radial velocities is still impractical, but measurement of the
radial velocity of the proper motion members identified here is feasible. This
will allow removal of a significant number of remaining interlopers \cite{Z99}
and will provide further information on the distribution of spectroscopic
binaries in these groups. Combined with the many new astrometric binaries
discovered by Hipparcos this will greatly improve our knowledge of the binary
population in OB associations.

The Hipparcos parallaxes are not sufficiently accurate to resolve the internal
structure of even the nearest associations. This would be interesting, as it
might help to delineate substructure, and hence shed light on the details of
the star formation process throughout an interstellar cloud. However, it is
possible to improve the individual distance estimates by using the proper
motions (and radial velocities) of established members to compute so-called
secular parallaxes (e.g., \cite{Dravins97}).

The lists of astrometrically selected members of OB associations are
incomplete beyond $V\sim7.3$, and a few genuine bright members (including some
long-period binaries) may have been excluded (see \cite{Z99} for details). The
lists extend to $V\sim10.5$, and include a few PMS objects in the nearest
groups. Available ground-based studies which can now be put on the Hipparcos
reference system can possibly complete the member list for the bright stars,
and extend them to fainter magnitudes. Unfortunately, the space motions of the
young groups are not large, and as a result the proper motions of the group
members do not differ very much from those of the field stars. Reliable
extension of the membership lists requires proper motions with accuracies of
order 2~mas~yr$^{-1}$ or better. These are provided by the ACT Catalog
\cite{Urban98} and the TRC Catalogue \cite{Hog98}, through a combination of
the positions in the Astrographic Catalog with the Tycho positions to
$V\sim11$. The individual stellar positions in these two catalogues have
modest accuracy, but the $\sim80$~yr epoch difference results in quite
accurate proper motions. The first installment of the ACT and TRC catalogues
both contain about 1 million measurements. The completion of the second
edition of the Tycho Catalogue \cite{Hog97} will provide proper motions of
similar quality for about 2.5 million objects to $V \sim 12$. However, the
1~mas accuracy parallaxes obtained by Hipparcos, which are generally not
available for fainter association members, play a crucial role in culling
interlopers from the membership lists. A complete study of all OB associations
in the Solar neighbourhood (extent, distance, structure, kinematics) has to
await the future GAIA space astrometry mission \cite{Perryman97},
\cite{Gilmore98}, which will gather astrometric data at the 10 micro-arcsecond
precision level to 20$^{\rm th}$ magnitude.

\subsection{Beyond the Galaxy}\label{sec:beyond}

Armed with a detailed understanding of associations in the Solar vicinity and
throughout the Milky Way, we can now venture further and investigate the vast
variety of extragalactic star forming regions, ranging from individual
associations to complete star bursts.

The nearest extragalactic associations are to be found in the Large Magellanic
Cloud. For example, at the heart of the 30~Doradus nebula one finds the
spectacular R136 cluster, which has a stellar density $\sim 200$ times greater
than that of a typical OB association and contains 39 identified O3 stars
\cite{Massey98b}.  This cluster is even considered a possible analogue of a
young globular cluster.  Thus by studying R136 we may gain a much more
detailed understanding of the `super star clusters' observed with the Hubble
Space Telescope in other galaxies (e.g., \cite{Conti94}, \cite{OConnel95},
\cite{Whitmore95}), which are also thought to be progenitors of globular
clusters.

One important lesson that has been learned from studying massive stars is that
because of their very high effective temperatures ($>30\,000$~K) it is vital
to perform spectroscopy of these stars if one wants to obtain reliable
physical parameters, such as masses \cite{Massey98a}. High sensitivity
combined with high spatial resolution are thus needed to perform detailed
studies of the luminous young star clusters observed in galaxies in the Local
Group and beyond. This will require the capabilities of the Hubble Space
Telescope or the largest ground-based facilities. A good understanding of
these extragalactic associations and the role they play will be very important
if we want to make sense of the observations of the earliest epochs of star
and galaxy formation that will eventually be carried out by future
millimeter-arrays and by the Next Generation Space Telescope.

%
%

\end{document}